\documentclass[journal,twoside,web]{ieeecolor}
\usepackage{generic}
\usepackage{cite}
\usepackage{amsmath,amssymb,amsfonts}
\usepackage{algorithmic}
\usepackage{graphicx}
\usepackage{textcomp}
\usepackage{color,soul}
\usepackage{pdfpages}

\usepackage{float}
\usepackage{fixltx2e}
\usepackage{subfig}
\usepackage{array, booktabs, makecell}

\sethlcolor{white} 
\DeclareRobustCommand{\hlyel}[1]{{\sethlcolor{white}\hl{#1}}} 

\def\BibTeX{{\rm B\kern-.05em{\sc i\kern-.025em b}\kern-.08em
	T\kern-.1667em\lower.7ex\hbox{E}\kern-.125emX}}
\begin{document}
\bstctlcite{IEEEexample:BSTcontrol}	
\title{Multispectral Video Fusion for\\ Non-contact Monitoring of\\ Respiratory Rate and Apnea}

\author{Gaetano Scebba, Giulia Da Poian, and Walter Karlen
	\thanks{This work was supported by the Swiss National Science Foundation (SNSF) under Grant 150640.}
	\thanks{G. Scebba, G. Da Poian and W. Karlen are with the Mobile Health Systems Lab, Institute of Robotics and Intelligent Systems, Department of Health Sciences and Technology, ETH Zurich, Switzerland (e-mails: gaetano.scebba@hest.ethz.ch, walter.karlen@ieee.org)}}

\maketitle

\begin{abstract}
	Continuous monitoring of respiratory activity is desirable in many clinical applications to detect respiratory events. Non-contact monitoring of respiration can be achieved with near- and far-infrared spectrum cameras. However, current technologies  are not sufficiently robust to be used in clinical applications.  For example, they fail to estimate an accurate  respiratory rate (RR) during apnea. We present a novel algorithm based on  multispectral data fusion that aims at estimating RR also during apnea. The algorithm independently addresses the RR  estimation  and  apnea  detection  tasks. Respiratory information is extracted from multiple sources and fed into an RR estimator and an apnea detector whose results are fused into a final respiratory activity estimation. We evaluated the  system retrospectively using data from 30 healthy adults who performed diverse controlled breathing tasks while lying supine in a dark room and  reproduced central and obstructive apneic events. Combining multiple respiratory information from multispectral cameras improved the root mean square error (RMSE) accuracy of the RR estimation from up to 4.64 monospectral data down to 1.60 breaths/min. The median F1 scores for classifying obstructive (0.75 to 0.86) and central apnea  (0.75 to 0.93) also improved. Furthermore, the independent consideration of apnea detection led to a more robust system (RMSE of 4.44 vs. 7.96 breaths/min). Our findings may represent a step towards the use of cameras for vital sign monitoring in medical applications.
\end{abstract}


\section{Introduction}
\label{sec:introduction}
\IEEEPARstart{M}{onitoring} respiratory activity is critical for assessing  the state of health  in humans. Respiratory rate (RR) is an important vital sign and a strong predictor of severe illness \cite{Cretikos2008}. While normal RR ranges between 12-20 breaths/min in healthy adults at rest,  RR outside of this range strongly correlates with specific adverse events, such as congestive heart failure \cite{flenady2017accurate}. 
Furthermore, monitoring respiratory activity is used to diagnose breathing disorders and lung diseases, such as pneumonia and central or obstructive sleep apnea. Despite its clinical relevance, the continuous and accurate monitoring of RR is highly undervalued \cite{Elliott2016}. 

\begin{figure}[!t]
	\centerline{\includegraphics[width=\columnwidth]{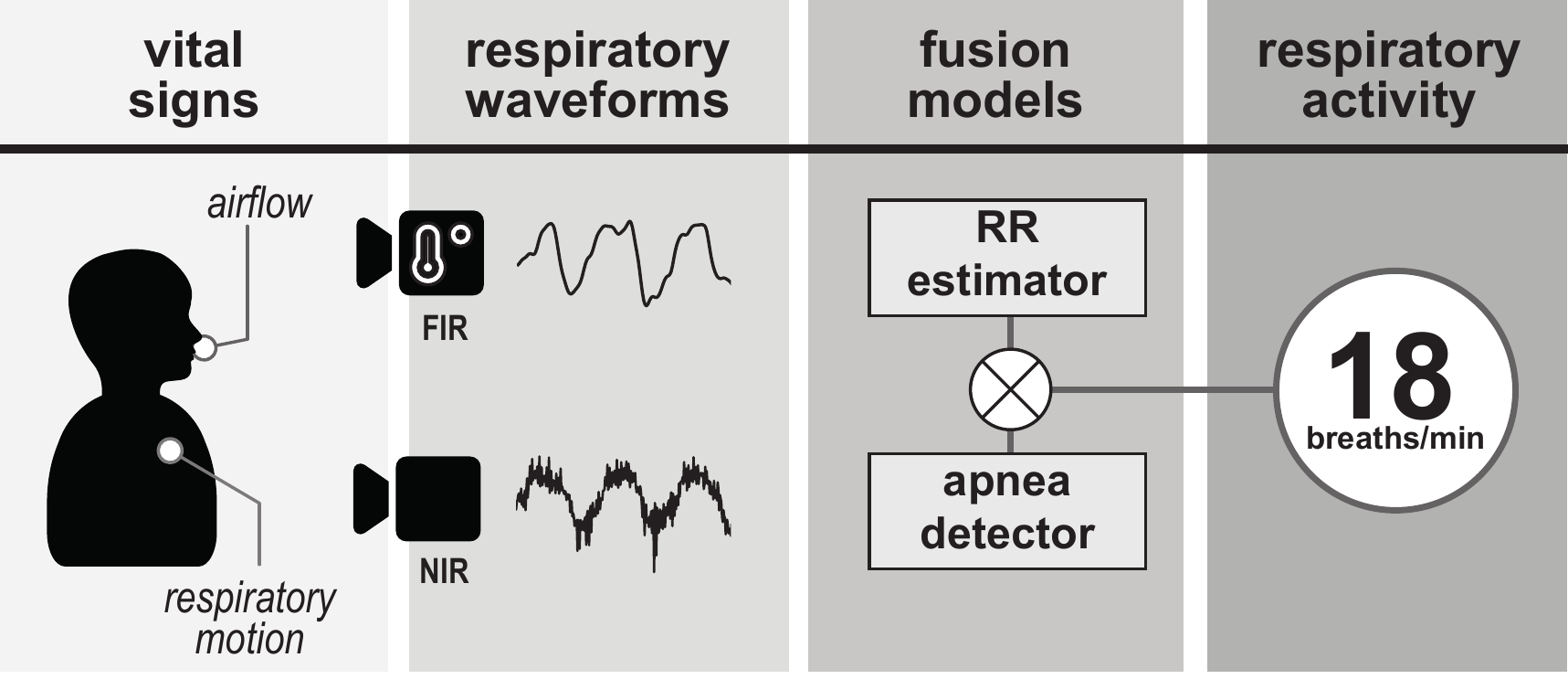}}
	\caption{Quantifying respiratory activity using far-infrared (FIR) and near-infrared (NIR) cameras. Dedicated fusion models exploit the respiratory information from the multispectral cameras and estimate respiratory rate (RR) and detect apnea.}
	\label{fig:teaser}
	\vspace{-0.4cm}
\end{figure}

Current approaches to objectively assess respiratory activity  require direct contact with the patient. The obtrusiveness of these approaches reduces their acceptability in the clinical and telemedicine settings \cite{Tarassenko2014}. 
Therefore, there has been an increased interest in developing nearables for non-contact,  unobtrusive monitoring of vital signs. One of these approaches includes the estimation of RR using cameras \cite{VanGastel2016a}. The ubiquitousness of cameras and the advancement in  computer vision enable physiological monitoring in more natural settings at home \hlyel{or in emergency settings for triage of potentially contagious patients} \cite{hegde2020autotriage}. Furthermore, the absence of direct contact with the skin allows for the monitoring of pre-term infants and other patients who suffer from skin irritations due to prolonged use of electrodes \cite{baharestani2007overview}. Another advantage of using cameras is the direct recognition of motion artifacts that reduce the vital sign quality and trigger false alarms \cite{muroi2019automated}.

Currently, cameras to monitor RR are  not widely adopted in clinical settings as they lack accuracy  during continuous operation.\hlyel{ The robustness of the measurement method is impacted in many ways.}  RGB cameras operating in the visible spectrum require moderate illumination, preventing their application in dark environments \cite{scebba2017improving}.  Most far-infrared (FIR) cameras are high-end products and excessively expensive, whereas consumer-accessible FIR cameras are challenged by low pixel resolution and sampling rates, which makes the localization of a specific region of interest (ROI) difficult \cite{Scebba2018}. An important limitation of most \hl{current} algorithms \hl{for camera-based respiration monitoring is} that they have only been tested under optimal conditions and with a low number of  subjects. Furthermore, many of these algorithms only implement a trivial apnea detection that fails under the presence of noise.

We propose a novel non-contact system to monitor respiratory activity by combining near-infrared (NIR) and FIR cameras  (Fig.~\ref{fig:teaser}). Our innovative approach extracts multiple streams of respiratory information from independent video modalities and then applies two dedicated data fusion models to determine whether the subject is breathing (apnea detection), and at which frequency (RR estimation). We designed and conducted a series of experiments to demonstrate that fusing multispectral videos leads to a significant improvement in the estimation of RR when compared to state-of-the-art methods, and increases the robustness of RR estimates when apnea events are present.





\section{Background}
\subsection{Multisensory fusion}

Data fusion has been successfully implemented in biomedical signal processing for RR estimation with conventional contact sensors \cite{charlton2017breathing}. Numerous approaches fuse either the RR from photoplethysmographic (PPG) \cite{Karlen2013a,pelaez2013,lazaro2013} or electrocardiographic (ECG) waveform modulations \cite{lazaro2014,clifford2010,orphanidou2013}. Others propose combining multiple sensor modalities, such as ECG and PPG \cite{lee2011,orphanidou2017,ding2016ptt}, ECG and thoracic impedance \cite{Shen2017impedance}, and ECG and accelerometry \cite{lepine2016}.  Fusing multiple vital signs is also advantageous in detecting apneic events. Furthermore, models exist that combine oxygen saturation with other modalities, such as ECG \cite{poupard2012,memis2017}, electroencephalography \cite{alvarez2009spectral}, respiratory effort and ECG \cite{al-angari2012}, nasal airflow \cite{sommermeyer2012} or tracheal sounds \cite{espinoza2015}. 

In this work, we build on  several of these approaches that have demonstrated the benefit of fusing multisensory information to increase accuracy and robustness of the RR estimation and apnea detection. However, while existing literature focused on vital signs obtained from contact sensors, our data fusion algorithms leverage respiratory signs that can be solely obtained from video data, thus enabling non-contact estimation of respiratory activity.

\begin{figure*}[t!]
	\centerline{\includegraphics[]{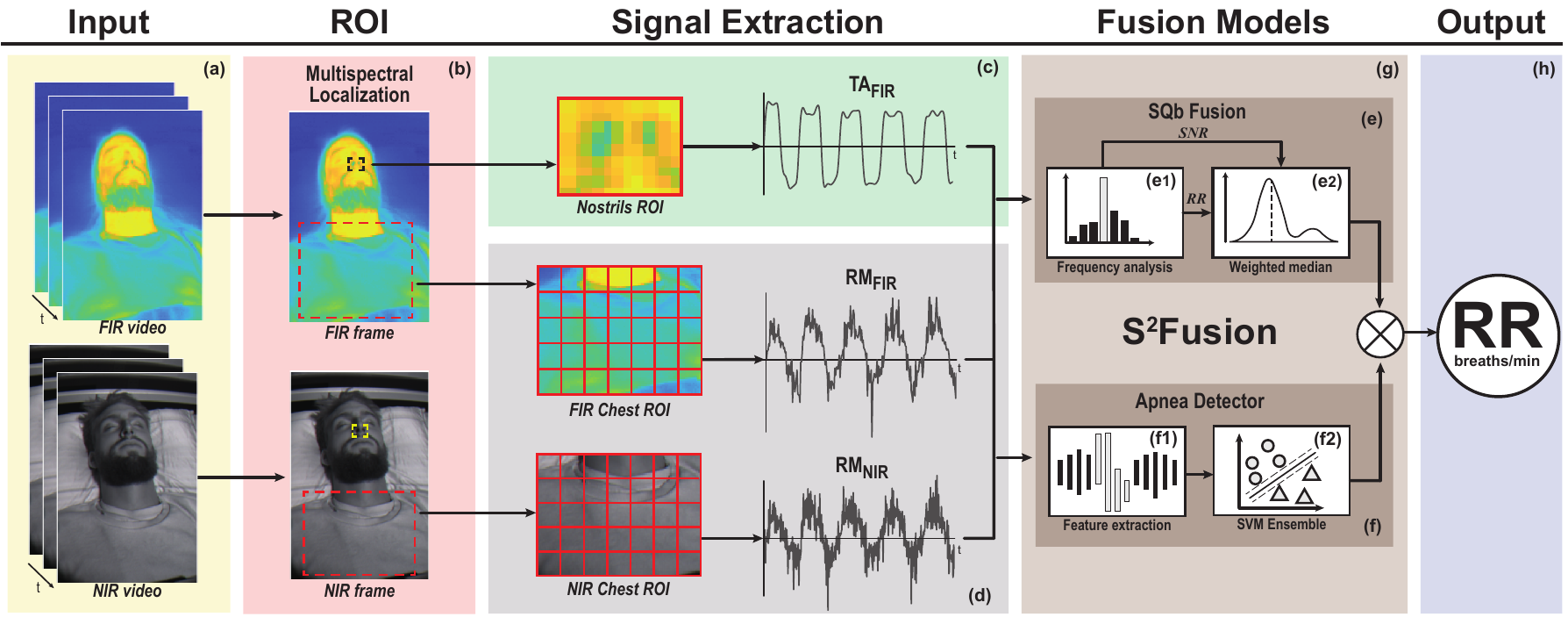}}
	\caption{Multispectral camera-based respiratory monitoring system. (a) The videos from far-infrared (FIR) and near-infrared (NIR) cameras are processed with a (b) Multispectral Region of Interest (ROI) localization algorithm \cite{Scebba2018}. The localized ROIs are used to extract (c) the Thermal Airflow (TA) signal from the nostril ROI, and (d) the Respiratory Motion signal both from the FIR and NIR chest ROIs. (e) In the Signal Quality based fusion (SQb Fusion), (e1) Respiratory Rate (RR) and Signal to Noise Ratio (SNR) are computed by the frequency analysis of the TA, and the  Respiratory Motion from the FIR (RM\textsubscript{FIR}) and the NIR  (RM\textsubscript{NIR}). (e2) The weighted median combines all RR estimates using their singnal-to-noise-ratio (SNR) as weights and calculates a  RR estimate. (f) The apnea detector (f1) extracts time and frequency features from the TA, RM\textsubscript{NIR}, and RM\textsubscript{FIR} signals and (f2) classifies the signals into apnea or breathing  with an ensemble of support vector machines. (g) The smart signal quality based fusion (S\textsuperscript{2}Fusion) fuses the results from the SQb Fusion and the apnea classifier (h) to obtain an apnea sensitive estimation of RR.}
	\label{fig:method}
\end{figure*}

\subsection{Respiratory activity detectable with digital cameras}
\label{sec:cam_signals}
Respiratory activity and its absence are characterized by specific patterns that can be extracted through appropriate analysis of videos recorded with RGB, NIR or FIR cameras. Four respiratory signs are directly observable: 1) respiratory induced motion, 2) thermal airflow variation, 3) respiratory plethysmographic modulation, and 4) apneic events.  

\subsubsection{Respiratory induced motion (RM)} RM of the torso is the most noticeable respiratory sign and is visible to the naked eye. It is produced by the volume variation of the lungs that generate the inflow and outflow of air. In particular, the lungs can be expanded and contracted generating the periodic motion of the torso in two distinct sections \cite{Guyton2006}. The motion of the abdominal wall is caused by the diaphragm that vertically lengthens and shortens  the lungs, and the elevation and depression of the rib cage increases and decreases the anteroposterior diameter of the lungs \cite{zordan2006breathe}. When an obstructive apneic event occurs, the collapse of the upper airways prevents air exchange, but a persistent respiratory movement of the torso is still present \cite{levy2015}. However, during a central apneic event, the RM is completely absent. Several algorithms have been proposed to quantify the RM. Differential image processing \cite{Bartula2013,wang2013unconstrained} or motion tracking of salient points extracted from the chest ROI with optical flow \cite{Li2017} and dense optical flow \cite{Janssen2016} enable the extraction of respiratory activity.   

\subsubsection{Thermal airflow (TA) signal} The temperature fluctuations of the airways due to airflow are a respiratory sign that can be detected with FIR cameras \cite{zhu2005tracking}. The physical phenomenon is based on the radiative and convective heat transfer component during the breathing cycle \cite{Abbas2011}, which results in a periodic increase and decrease of the temperature at the tissues around the nasal cavity. These observable temperature fluctuations are quantifiable in a FIR video as pixel intensity variations of the nostrils ROI \cite{Abbas2011,Pereira2015,Cho2017}. During obstructive and central apnea there is a suspension of the air exchange between lungs and atmosphere and consequently, no temperature variation is observable at the nostrils. 

\subsubsection{Respiratory plethysmographic (RP)  modulation} The various respiratory modulations of the PPG waveform are well described in literature \cite{Karlen2013a}. The most prominent respiratory sign observable with cameras is the PPG baseline modulation. Each respiratory cycle causes variations of the intra-thoracic pressure that induce changes in the blood exchange between the pulmonary and systemic circulation. This phenomenon results in a periodic variation of the baseline of the PPG waveform. In the event of obstructive apnea, the persistent motion of the torso continues to induce the intrathoracic pressure variations. During a central apnea event, the RP modulation is drastically reduced in amplitude due to the absence of any RM. The PPG baseline modulation can be quantified through the linear combination of the RGB components in a video of skin \cite{VanGastel2016a}. Other studies apply blind-source separation \cite{Poh2011,Jorge2018}, continuous wavelet filtering \cite{Bousefsaf2013}, or auto-regressive models \cite{Tarassenko2014} to extract the RP modulation.

\subsubsection{Apneic events} Apneic events are defined as the suspension of respiratory activity lasting for more than 10 seconds \cite{Berry2013}. Apnea detection algorithms frequently focus on detecting an absence of RR \cite{VanGastel2016a,pereira2018noncontact}. However, this is not reliable as frequency derived RR does not tend towards zero if there is noise present. Furthermore, this simplistic view on apnea detection  prevents the  distinction between the absence of breathing and the absence of a respiratory signal, as well as between obstructive and central apnea.

Pereira \textit{et al.} have exploited the potential of combining multiple respiratory sources of information obtained from cameras \cite{Pereira2016}. They propose a black-box  fusion approach, in which the TA fluctuations and the RM obtained from high-end FIR cameras are considered as independent data sources \cite{pereira2018noncontact}. 


In contrast to the above described work, we present the first camera-based approach to monitor a broader range of respiratory activity by combining multiple types of respiratory modulations extracted from multiple modalities. Our approach simultaneously extracts multiple respiratory signals and with specialized data fusion models estimates the RR and classifies apnea. \hlyel{With a third fusion stage we combine that information to produce an  apnea-aware RR quantification.}

\section{Algorithm development}
We developed a system for estimating respiratory activity from synchronized NIR and FIR cameras (Fig.~\ref{fig:method}). We chose this pair as they can be operated independently of visible light. Our algorithmic approach consisted of three main steps. Firstly, we localized the nostril and chest ROIs in both the NIR and FIR spectra. Secondly, three respiratory signals were extracted from these ROIs. 
 Finally, using our novel Smart Signal quality based Fusion (S\textsuperscript{2}Fusion), a RR estimate obtained from a modulation fusion model and   an apnea classification model were combined to obtain a more accurate quantification of the respiratory activity.

\subsection{Multispectral ROI localization}
The detection and tracking of the ROIs  was built upon a multispectral ROI detector that we previously proposed in \cite{Scebba2018}. 

\subsubsection{Image registration}
In order to ensure the pixel-to-pixel correlation between the FIR and NIR frames, we applied a spatial image registration model, consisting of an affine transformation $T$, such as
\begin{equation}
T =
\begin{bmatrix}
s_x & 0 & t_x \\
0 & s_y & t_y\\
\end{bmatrix},
\label{eq:registration}
\end{equation}
where $s_x$ and $s_y$  specified  the scaling factors and $t_x$ and $t_y$ the translation factors. The definition of these factors was dependent on the distance between the object of interest and the camera \cite{Scebba2018}. Therefore, we defined the transformation model during the calibration phase of the cameras and the same factors were applied for all the recordings. The spatial alignment between FIR and NIR frames enabled the projection of the ROIs identified in the NIR frame to the FIR frame.

\subsubsection{Detection and tracking}
Five facial landmarks were detected in the NIR frame by applying a cascade convolutional neural network (CCNN) \cite{Zhang2016}. We obtained the nostril ROI as a rectangle centered over the nose landmark, with 20~pixel height and 15~pixel width. To locate the chest ROI, we used the coordinates of the chin landmark, which were derived as the bottom ordinate of the face box obtained from the CCNN model. The final chest ROI was defined using frame size dependent geometrical relations.

After successful ROI detection, an object tracker compensated for the movements of the subjects during the recording. As the chest ROI was defined by geometrical dependencies between the nostrils and face position, we tracked the nostril ROI on the NIR video only. 
The tracker first extracted feature points from the ROI using the minimum eigenvalue algorithm \cite{Shi1994}, then tracked these points with the Kanade-Lucas-Tomasi single points tracker \cite{klt}, followed by correcting potential tracking failures with the method described by Kalal \textit{et al.} \cite{kalal2010}. To further mitigate potential tracking errors, we re-triggered the detection of the nostril ROI every 10 seconds.

\subsection{Signal extraction}
We extracted the RM from NIR and FIR and the TA from FIR videos. As a preprocessing step, we applied median filtering with a 3$\times$3 mask to each NIR frame, and we enhanced the contrast in each FIR frame by normalizing the pixel value to the interval $[0,1]$. 
Following the preprocessing, we extracted the respiratory signals from each of the ROIs. The signals were then filtered and the RR was extracted from the maximal power spectral density.  

\subsubsection{Thermal airflow}
The thermal fluctuations at the nostril ROI resulted in a pixel intensity variation in the FIR video. Thus, given a correctly identified ROI at time $t$, we computed the TA\textsubscript{FIR} signal as the spatial average of the pixel intensity within the nostril ROI (Fig.~\ref{fig:method} c) such as,
\begin{equation}
TA_{\text{FIR}}(t) = \frac{1}{WH}\sum_{x}^{W}\sum_{y}^{H}I_{\text{ROI}}(x,y,t),
\label{eq:TA}
\end{equation}
where $W$ and $H$ are the width and height of the ROI, and $I_{\text{ROI}}(x,y,t)$ is the pixel intensity at pixel $x$, $y$ at time $t$.

\subsubsection{Respiratory induced motion}
The periodic motion of the torso was extracted using Farnebäck dense optical flow algorithm \cite{farneback2003two}. \hl{In particular, we used the vertical velocity profiles of each tracked pixel, which were obtained as the ratio between the pixel displacement derived by the optical flow algorithm and the time between two consecutive frames. In contrast to estimating the optical flow for all the pixels of each frame} \cite{Li2017}, \hl{we restricted the optical flow calculation to the pixels within the chest ROI only.}  For this, we divided the ROI into a 5$\times$7 cell grid and averaged the velocity profiles within each cell, obtaining 35 velocity profiles (Fig.~\ref{fig:method} d). To reduce the number of the velocity profiles for further processing, we calculated the standard deviation $\sigma_{p}$ of each profile over a window of 12 s and then excluded the profiles that did not meet the criterion, such as
\begin{equation}
\sigma_{p} < \text{median}(\mathbf{\Sigma_{p}}) + 2\cdot \text{IQR}(\mathbf{\Sigma_{p}}), 
\label{eq:vel_profiles}
\end{equation}
where $\Sigma_p$ is a vector containing the standard deviation of the 35 velocity profiles and IQR was the interquartile range. The RM signal was obtained by averaging the remaining velocity profiles. As the multispectral ROI localization enabled the identification of the chest ROI in the NIR and FIR frames, we extracted a RM signal from each spectral video, thus obtaining the RM\textsubscript{NIR} and RM\textsubscript{FIR} signals (Fig.~\ref{fig:method} d).

\subsubsection{RR estimation}
The TA\textsubscript{FIR}, RM\textsubscript{NIR}, and RM\textsubscript{FIR} signals were processed to estimate the respective RR values. 
For each signal, we applied the regularized least squares detrending algorithm introduced by Tarvainen \textit{et al.} \cite{tarvainen2002advanced}, with the smoothing parameter $\lambda=300$. In the case of the TA signals, we also applied a Butterworth band pass filter, with filter order set to 2 and frequency range to 0.015 – 0.75~Hz. We did not apply any filter to the RM\textsubscript{NIR} and RM\textsubscript{FIR} signals to preserve their information content for the apneic detection task. The signals were then windowed and the Lomb-Scargle power spectral density (PSD) \cite{lomb1976least,scargle1982studies} computed. The frequency with the highest power density in the range of 0.015 – 0.75~Hz was selected and converted to breaths/min. The Lomb-Scargle PSD estimator was chosen because cameras often have unstable frame rates and it did not require an evenly sampled time series. 

\subsection{Smart signal quality based fusion (S\textsuperscript{2}Fusion)}
The S\textsuperscript{2}Fusion was designed as a multilevel data fusion algorithm that processes respiratory signals extracted from independent multispectral videos in order to quantify the respiratory activity. Practically, it consisted of three components: 1) a signal quality based fusion (SQb Fusion) to merge the RR estimates computed from each respiratory signal, 2) an apnea detector to extract temporal and spectral features from the respiratory signals and classify them as either an apnea or respiratory epoch, and 3) a final fusion to combine both models into a respiratory activity estimation (Fig.~\ref{fig:method} g).

\subsubsection{Signal quality based fusion}
The SQb Fusion algorithm estimated the RR considering the signal quality from each of the three respiratory signals (TA\textsubscript{FIR}, RM\textsubscript{NIR}, and RM\textsubscript{FIR}).  The signal-to-noise ratio 
was computed 
such as 
\begin{equation}
\text{SNR}_{i} = \left(\frac{ \sum_{f_{\text{peak}} - \frac{k}{2}}^{f_{\text{peak}} + \frac{k}{2}} P(f)}{\sum_{0}^{f_{\text{peak}}-\frac{k}{2}} P(f) +\sum_{f_{\text{peak} +\frac{k}{2}}}^{f_{\text{s}}} P(f)}\right)_i,
\label{eq:snr}
\end{equation}
where $P(f)$ was the Lomb-Scargle PSD, $f_{\text{peak}}$ was the frequency of the peak with the highest power density, $k~=~2$~breaths/min was the margin parameter \cite{DeHaan2013}, and $f_{\text{s}}$ the sampling frequency. RR\textsubscript{\text{SQb}} was obtained from the weighted median using the RR\textsubscript{i} from each signal $i$ and the weights $w_i$  such as
\begin{equation}
w_i = \frac{\text{SNR}_i}{\sum_i \text{SNR}_i} \cdot M,
\label{eq:sqb_weights}	
\end{equation}
where $M = 24$ was an empirically determined scaling factor.


\subsubsection{Apnea detector}
We separated  respiratory epochs from apneic epochs with an ensemble of support vector machines (SVMs, Fig. S1, Supplementary Material).\hlyel{ In addition to the SNR used for the SQb fusion,  we extracted time domain features such as the number of mean crossings,  variance, and standard deviation  from the 12 s windows}. These features highlighted the morphological changes of the signals during the apneic events. In contrast, the frequency domain feature SNR described the periodicity of the breathing activity. \hlyel{All features were standardized to have zero mean and unit variance before they were fed into the   SVM ensemble classifier.}
\hlyel{Test features were standardized based on training data statistics.}
The SVM consisted of one layer of  regression models $R_i$ 
and one classification model $C$ (Fig. S1, supplementary material). Each $R_i$ extracted the posterior probability $Pr(Y = apnea|X_i)$ for the set features $X_i$ to belong to an apneic epoch. This way, each  $R_i$ consisted of an expert model for each respiratory signal. The purpose of $C$ was to aggregate the $Pr$ from each expert $R_i$ and to compute the final output y\textsuperscript{apnea}. The design of the SVM ensemble was inspired by the work of Schwab \textit{et al.} on multitasking networks \cite{Schwab2018b}.

\section{Methods and materials}
We conducted a series of experiments to test whether 1) the SQb Fusion algorithm is more accurate in estimating the RR compared to a monospectral approach, 2) the combination of respiratory sources extracted from independent multispectral cameras increases the detection rate of apneic events, 3) 
the S\textsuperscript{2}\hlyel{Fusion algorithm produces robust RR during apneic events and} can substantially improve the accuracy during recordings that include central or obstructive apneic events, \hlyel{and 4) there is a performance bias towards a subject's sex.}

\subsection{Experimental protocol}
To evaluate the performance of our video-based respiratory monitoring algorithm, we designed an experimental protocol consisting of one task where participants were breathing spontaneously and four tasks where they followed different breathing patterns that included apneic events. Participants lay on a bed in a supine position and for each task breathed through the nose. During the tasks with apneic events, they breathed following the rhythm of a metronome. We only considered the supine body posture because it is the most common position leading to apneic events during sleep 
\cite{white2005pathogenesis}. To challenge 
our algorithms, all videos were recorded in a dark environment ($<$ 5 Lux). 

\begin{figure}[t]
	\includegraphics[width=\columnwidth]{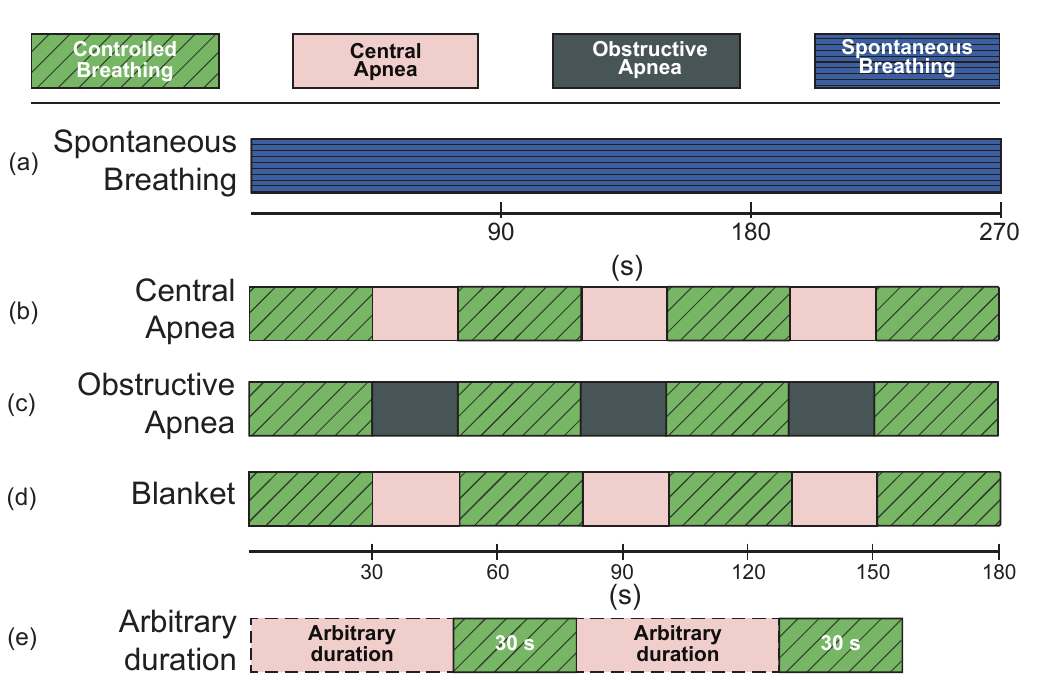}
	\caption{The experimental protocol consisted of 5 different tasks:  (a) spontaneous breathing for 4.5 minutes; constant breathing at 10~breaths/min interrupted by apneic events that simulated (b) central, (c) obstructive, (d) central  while the body was covered with a blanket, and (e) central with arbitrary duration based on subject breath holding capacity.}
	\label{fig:protocol}
\end{figure}

\textbf{Spontaneous Breathing} Participants breathed spontaneously for 4.5 min. During this task, participants turned their head  45\textsuperscript{o} left and right  for 2 min each (Fig.~\ref{fig:protocol}~a).

\textbf{Central Apnea} Participants breathed for 3 min with a constant RR of 10~breaths/min and performed 3 central apneic events, each lasting 20 s (Fig.~\ref{fig:protocol}~b).

\textbf{Obstructive Apnea} Participants breathed for 3 mins with a constant RR of 10~breaths/min and performed 3 obstructive apneic events, each lasting 20 s. For each apneic event occurred, participants were instructed to simulate a blockage of the airways, and to keep the thorax moving to mimic an ongoing respiratory effort (Fig.~\ref{fig:protocol}~c).

\textbf{Central Apnea - Blanket} Participants breathed for 3 min with a constant RR of 10 breaths/min and performed 3 central apneic events, each lasting 20 s. During the entire duration of this task, participants were covered to the chin with a blanket to hide the chest contours (Fig.~\ref{fig:protocol}~d).

\textbf{Central Apnea - Arbitrary duration} Participants breathed with a constant RR of 10~breaths/min and 2 central apneic events whose duration was based on the participant’s breath holding capacity (Fig.~\ref{fig:protocol}~e).

The study was conducted according to the ethical guidelines of Helsinki. Institutional research ethics board approval was obtained (ETH Zurich EK 2017-N-60). After informed written consent, healthy volunteers with no history of cardio-respiratory disease were enrolled in the study.  

\subsection{Experimental setup}
The setup consisted of two cameras interfaced with a Raspberry Pi 3 B microcomputer (Raspberry Pi Foundation, Cambridge, UK). Custom software was developed to ensure simultaneous recording of the cameras, video compression, and data storing on the SD card. NIR videos were recorded with the See3cam\_CU40 (econ-Systems, Chennai, India) with a frame rate of 15 Hz and a resolution of 336$\times$190 pixels. FIR videos were recorded with a FLIR Lepton version 3.5 (FLIR Systems Inc., Wilsonville, United States) with an average frame rate of 8.7 Hz and a resolution of 160$\times$120 pixels. We used a NIR LED array to enable recordings in insufficient lighting.

Reference respiratory effort was recorded using a certified ezRIP module (Philips Respironics, Pennsylvania, USA) at a sampling rate of 50 Hz. The device used two \hl{elastic bands equipped with piezo-resistive sensors} placed over the thorax and abdomen\hl{, which is one of the recommended setups for measuring respiratory effort in clinical polysomnography} \cite{Berry2013}.

Collected videos and reference respiratory signals were transferred to a PC, where additional processing and analysis were performed with Matlab R2018b (MathWorks Inc., Natick, USA). The sliding window size was set to 12 s for all processing steps and was recomputed at the lowest camera sampling rate (FIR, \texttildelow8.7 Hz).

\subsection{SVM ensemble training and testing}
\label{sec:svm_training}
The adoption of the SVM ensemble for apnea detection implied a split of the dataset that could ensure a fair evaluation of the model. For this reason, we applied the \hl{leave-one-subject-out} validation scheme, \hl{which is a particular case of the K-fold validation with K equal to the number of subjects. This validation scheme allowed us to maximize the amount of data used for training while accounting for potential high variance problems} \cite{kohavi1995study}. At each run, the training and test sets included the data of N-1 subjects and 1 subject respectively, with N equal to the number of recruited subjects. The training set consisted of the recordings obtained from \textit{Central Apnea} and \textit{Obstructive Apnea}, whereas the test set included the recordings obtained from all the tasks with apneic events.  We did not include the recordings obtained from the \hl{\textit{Spontaneous Breathing},}  \textit{Central Apnea – arbitrary duration}, and \textit{Central Apnea – covered with blanket} tasks in the training set because they did not provide additional information regarding the signal morphology of apneic events \hl{and would have increased the class imbalance} (Fig. S2, Supplementary Material). 

\subsection{Evaluation}

The performance of the SQb Fusion algorithm for RR estimation was compared against algorithms that estimate the RR from single respiratory signals before fusion. 
Additionally, we implemented two baseline fusion algorithms based on mean and median of the RR estimates obtained from RM\textsubscript{NIR}, RM\textsubscript{FIR}, and TA\textsubscript{FIR} signals. The RR estimates of all the evaluated algorithms were reported independently of their underlying estimation quality. 

The performance of the SVM ensemble for apnea detection, which was trained with the features extracted from all signals (RM\textsubscript{NIR}+RM\textsubscript{FIR}+TA\textsubscript{FIR}), was compared to the performance of three baseline SVM binary classifiers, trained with the features extracted from the individual RM\textsubscript{NIR}, RM\textsubscript{FIR}, and TA\textsubscript{FIR} signals (Fig. S3, Supplementary Materials). 

We evaluated the performance of the S\textsuperscript{2}Fusion to quantify the respiratory activity on a combined experiment that includes both the RR estimation and apnea detection tasks. In particular, we compared it against the SQb Fusion using all the recordings that included apneic events. To guarantee an objective evaluation without overlap \hl{between} training and test set \hl{in the S\textsuperscript{2}Fusion evaluation experiment}, we applied the \hl{same leave-one-subject-out} validation scheme introduced in section \ref{sec:svm_training}.

\subsection{Performance metrics}
\hl{In order to describe the performance of the evaluated algorithms, we evaluated the accuracy (agreement between the RR estimates and a reference) and robustness (accuracy distribution in challenging or altering situations, specifically apneic events or differences in demographics).}

To obtain accuracy, we calculated Bland-Altman plots, Pearson's correlation coefficient, and the root mean square error (RMSE). 
As reference RR, we used the RR estimates obtained from the \hlyel{thorax} respiratory effort sensor. For comparison, we pooled all the estimated RRs within non-overlapping segments of 15 s and computed the median thereof.
Bias and limits of agreement (LoA) for Bland-Altman analysis were calculated using the formulas for repeated measurements \cite{Bland1999}. Pearson's correlation coefficient $r$ was computed using a 95\% confidence interval. The RMSE was computed between $RR_{\text{ref}}$ and each fusion algorithm \hlyel{across segments for each subject and overall.} 
\hlyel{F1 score, sensitivity, and specificity were calculated to compare the apnea classification.}

\hlyel{To describe robustness, across-subject RMSE distributions were split by sex and displayed as boxplots. Also, we compared the across-subject RMSE distributions between the RR estimates obtained from the S\textsuperscript{2}Fusion and the SQb Fusion algorithms to evaluate each apneic event task.}    Middle, bottom, and top horizontal lines of boxes depicted the  median, lower, and upper quartile, respectively, and crosses depicted outliers. 

We tested the normality of the RMSE distributions using the Shapiro-Wilk test and compared them with the Wilcoxon Rank Sum test.  Multiple distribution comparisons were Bonferroni corrected. Significance levels were set to $p<0.001$ (***) or $p<0.05$ (*) unless stated otherwise. 


\section{Results}
\begin{figure}[t!]
	\centering
	\includegraphics[width=\columnwidth]{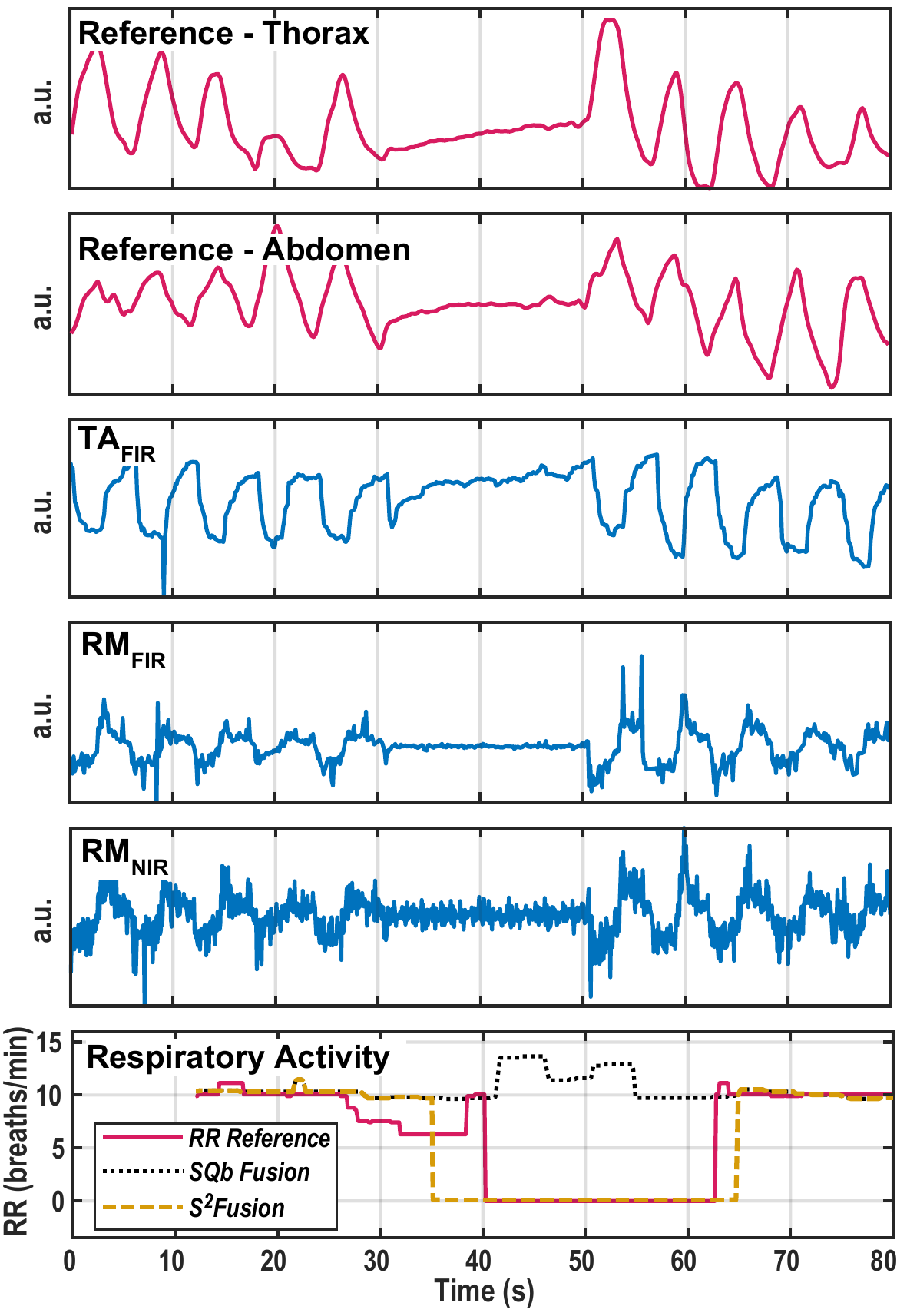}
	\caption{Comparison of respiratory signals from one subject during a central apnea task for (from top) reference respiratory effort \hlyel{ from the thorax, the  abdomen},  TA\textsubscript{FIR},  RM\textsubscript{FIR},  RM\textsubscript{NIR}  signals, \hlyel{and respiratory activity (RA) estimations.}  SQb Fusion and S\textsuperscript{2}Fusion algorithm were only available after the first 12 s window was collected.}
	\label{fig:showcase}
	\vspace{-0.5cm}
\end{figure}

We obtained videos of 30 healthy participants (17 females and 13 males, mean age: 27~$\pm$~3 y). Subjects varied in facial appearance with a Fitzpatrick skin tone between type I and VI  (I:2, II:7, III:19, IV:1, V:1, VI:0). 
Seven  males had facial hair. A total of 492 minutes of video were recorded and available for analysis. An illustrative example of the signals obtained from the reference contact sensors and cameras, as well as estimation output, is depicted in Fig~\ref{fig:showcase}. 
  Motion artifacts substantially corrupted the recording of one subject. 
  
\begin{table}[th!]
\caption{Apnea detection performance for all apnea tasks on the test set. Median (mean absolute deviation) for F1 score, sensitivity, and specificity were obtained for the SVM ensemble trained with all the features,  and features from the individual respiratory signals RM\textsubscript{NIR}, RM\textsubscript{FIR} and TA\textsubscript{FIR}. 
} 	
\centering	
\begin{small}
	\begin{tabular}{llll}
		\toprule
		\multicolumn{4}{c}{\textbf{Obstructive Apnea}}\\
		\midrule
		Source & F1 & Sensitivity & Specificity  \\			
		\midrule
		RM\textsubscript{NIR}+RM\textsubscript{FIR}+TA\textsubscript{FIR} & \textbf{{0.86}} (0.12) & \textbf{{0.91}} (0.16) & {0.78} (0.06)\\ 			  
		RM\textsubscript{NIR} & {0.76} (0.12) & {0.75} (0.17) & \textbf{{0.82}} (0.05)\\ 				
		RM\textsubscript{FIR} & {0.76} (0.15) & {0.80} (0.20) & {0.69} (0.05)\\  				
		TA\textsubscript{FIR} & {0.75} (0.14) & {0.74} (0.17) & {0.72} (0.08)\\ 						 			 			
		\midrule
		\multicolumn{4}{c}{\textbf{Central Apnea}}\\
		\midrule
		RM\textsubscript{NIR}+RM\textsubscript{FIR}+TA\textsubscript{FIR} & \textbf{{0.93}} (0.02) & \textbf{{0.98}} (0.03) & {0.86} (0.04)\\   
		RM\textsubscript{NIR} & \textbf{{0.93}} (0.08) & {0.95} (0.13) & \textbf{{0.92}} (0.06)\\
		RM\textsubscript{FIR} & {0.88} (0.04) & \textbf{{0.98}} (0.06) & {0.81} (0.06)\\ 			  					
		TA\textsubscript{FIR} & {0.75} (0.10) & {0.77} (0.13) & {0.70} (0.07)\\    			
		\midrule
			\multicolumn{4}{c}{\textbf{Central Apnea - Arbitrary duration}}\\
		\midrule
		RM\textsubscript{NIR}+RM\textsubscript{FIR}+TA\textsubscript{FIR}& {0.87} (0.07) &\textbf{{0.96}} (0.10) & {0.82} (0.08) \\   
		RM\textsubscript{NIR} & \textbf{{0.89}} (0.08) & {0.91} (0.09) &\textbf{{0.87}} (0.08)\\
		RM\textsubscript{FIR} & {0.87} (0.08) & {0.89} (0.11) & {0.86} (0.07) \\ 			  					
		TA\textsubscript{FIR} & {0.72} (0.12) & {0.69} (0.14) & {0.78} (0.08)\\    		
		\midrule
			\multicolumn{4}{c}{\textbf{Central Apnea - Blanket}}\\
		\midrule
		RM\textsubscript{NIR}+RM\textsubscript{FIR}+TA\textsubscript{FIR} & {0.92} (0.03) & \textbf{{1.00}} (0.09) & {0.86} (0.05)\\   
		RM\textsubscript{NIR} & \textbf{{0.94}} (0.04) & {0.98} (0.10) &\textbf{{0.92}} (0.07)\\
		RM\textsubscript{FIR}& {0.90} (0.05) & {0.99} (0.11) & {0.81} (0.08)\\ 			  					
		TA\textsubscript{FIR} & {0.74} (0.15) & {0.79} (0.19) & {0.72} (0.07)\\    		
		\toprule
	\end{tabular}
	\label{tab:all}
\end{small}
\end{table}

\begin{figure*}[t!]
	\centering
    \includegraphics[width=\linewidth]{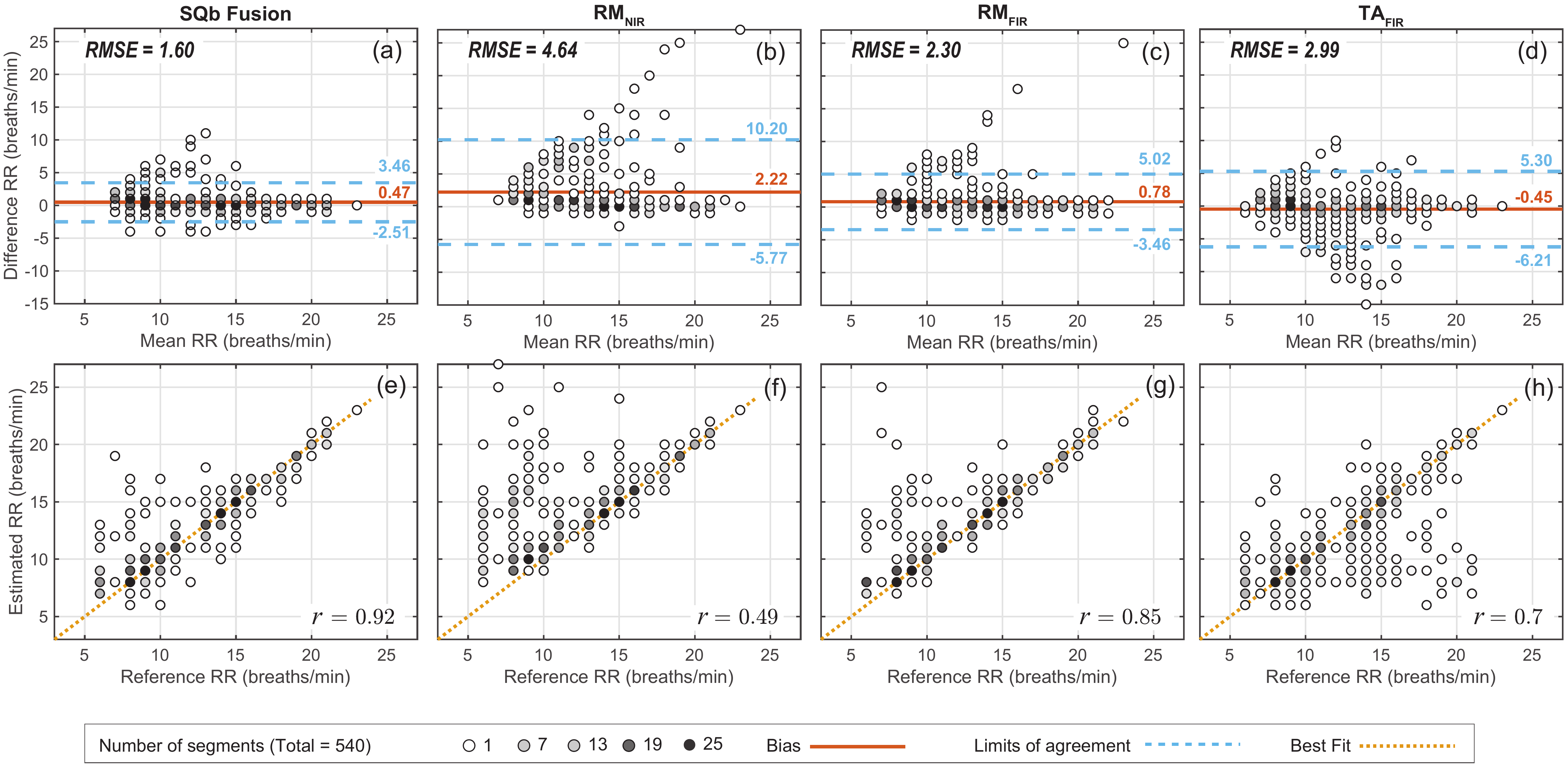}
	\caption{Bland-Altman (a-d) and scatter plots (e-h) of the estimated RR using (a, e) the Signal Quality based (SQb) fusion and the individual respiratory signals (b, f) RM\textsubscript{NIR}, (c, g) RM\textsubscript{FIR}, and (d, h) TA\textsubscript{FIR} on the Spontaneous Breathing data divided into 15 s long segments. The dashed blue lines depict the limits of agreement, the solid red line the bias. The dotted yellow line depicts the linear relation that best fits the data. $r$ is the Pearson's correlation coefficient and RMSE the root mean square error across all segments.}
	\label{fig:ba}
\end{figure*}

\subsection{Respiratory rate estimation performance}
\begin{figure}[t]
	\centering
	\includegraphics[width=\columnwidth]{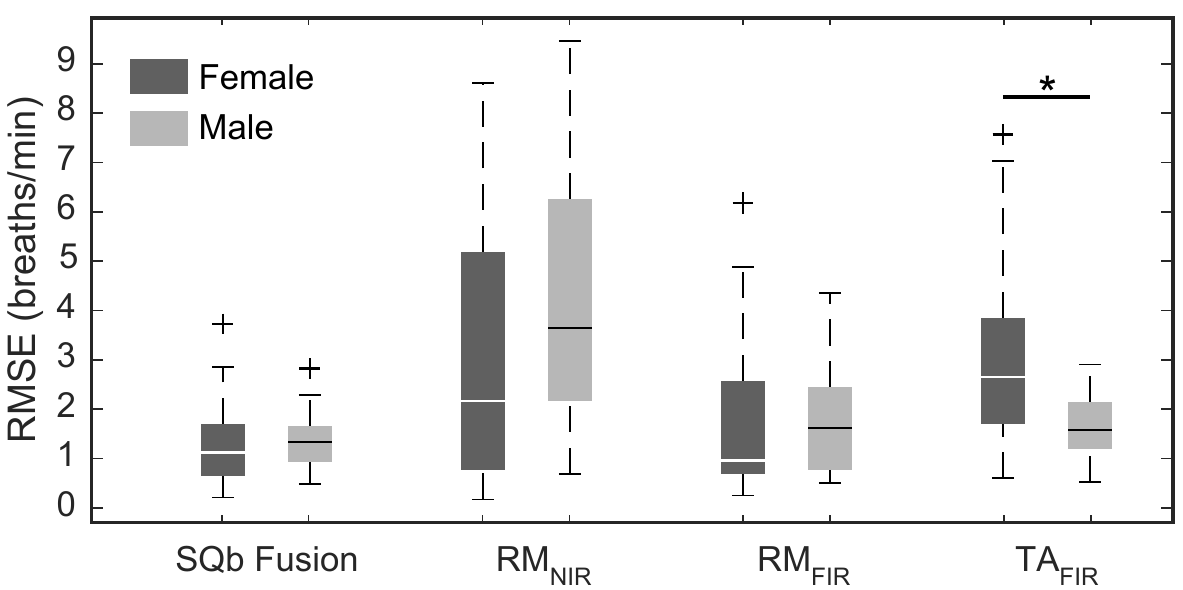} 
	\caption{Boxplot comparison of the root mean square error (RMSE) from the Spontaneous Breathing data between subjects of different sex for the signal quality based (SQb) Fusion algorithm and the algorithms based on the individual respiratory signals. 
	}
	\label{fig:genderAnalysis}
	\vspace{-0.3cm}
\end{figure}

A total of 135 minutes (540 segments) of \textit{Spontaneous Breathing} task were evaluated for the RR estimation \hlyel{accuracy}. The SQb fusion showed the highest accuracy (Fig.~\ref{fig:ba}). 
This was supported by the highest correlation ($r$=0.92), lowest RMSE (1.6~\hl{breaths/min}), and the best agreement from the Bland-Altman analysis. The bias was 0.47~breaths/min and the LoA were 3.46 and -2.51~breaths/min, outperforming all the algorithms based on individual respiratory signals from single video modalities (Fig.~\ref{fig:ba}~e-h). 
The SQb Fusion obtained a median RMSE of 1.17~breaths/min across subjects, which was significantly lower than the baselines (median fusion 1.60~breaths/min, $p<0.05$ and mean fusion 1.96~breaths/min, $p<0.01$, Table S1, Fig. S4, Supplementary Materials). 
The RMSE of TA\textsubscript{FIR} showed a significantly different distribution between female and male subjects (median RMSE 2.65~vs.~1.58 breaths/min, Fig.~\ref{fig:genderAnalysis}).

\subsection{Apnea detection}
A total of 357 minutes of video recordings containing apnea were evaluated for apnea and respiratory activity detection. Fusing the features extracted from multispectral videos led to the best performance in detecting apneic events in the \textit{Obstructive Apnea} task (F1 score = 0.86, Table~\ref{tab:all}). 
The performance of the SVM ensemble trained with all the features or using only those extracted from the RM\textsubscript{NIR} signal showed similar results on the \textit{Central Apnea} task (F1 score = 0.93 for both sources).


\subsection{Respiratory activity detection}
\begin{figure}[t!]
	\centering
	\includegraphics[width=\columnwidth]{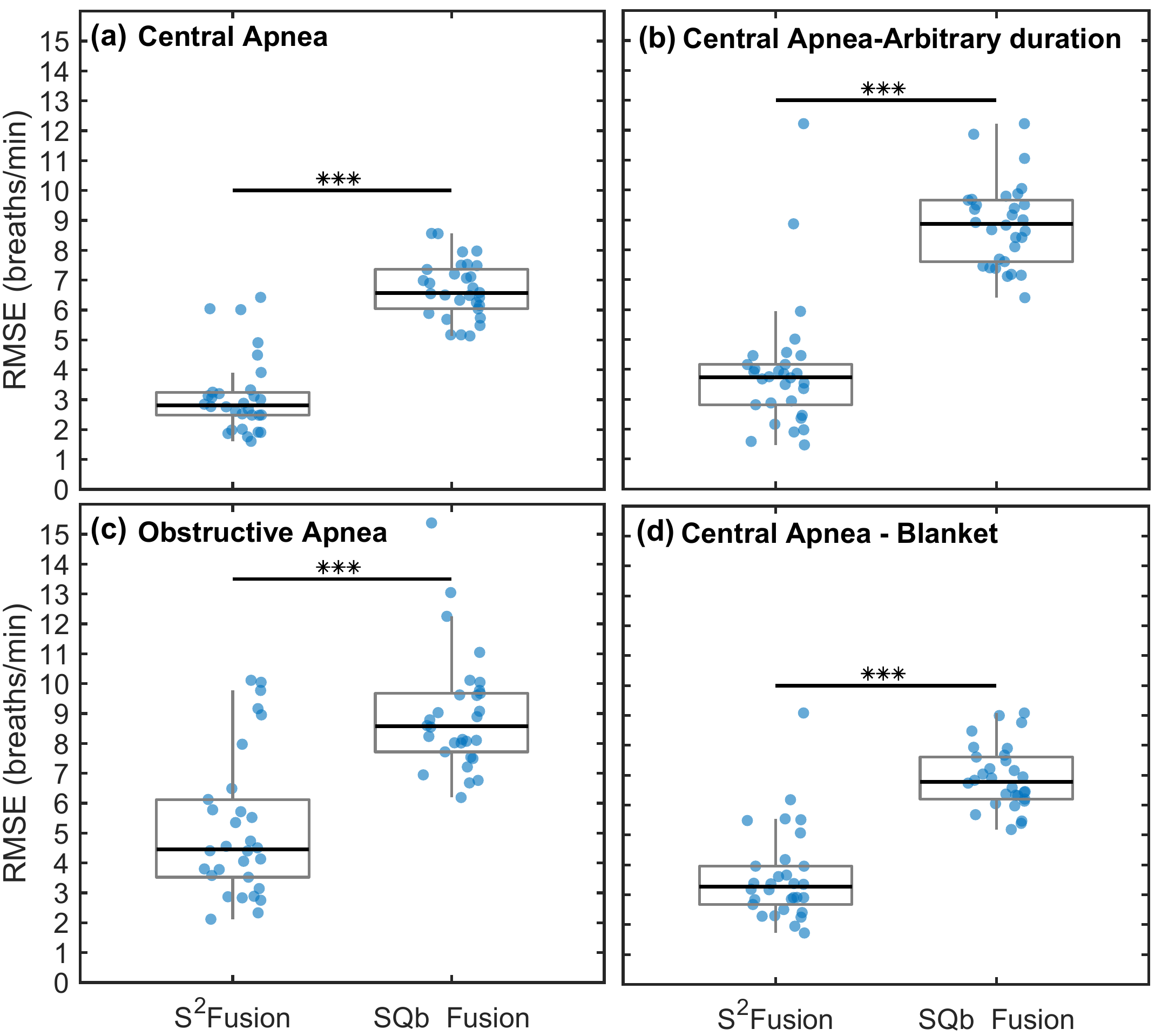}
	\caption{Comparison of the root mean square error (RMSE) for the S\textsuperscript{2}Fusion and SQb Fusion algorithms by apneic event type: (a) Central Apnea, (b) Central Apnea – Arbitrary duration, (c) Obstructive Apnea, (d) Central Apnea – Blanket. 
	} 
	\vspace{-0.3cm}
	\label{fig:bplots}	
\end{figure}

The S\textsuperscript{2}Fusion algorithm showed a statistically significant reduction of RMSE compared to the SQb Fusion in all apnea tasks (Fig.~\ref{fig:bplots}). \hl{The RMSE over all tasks was 4.44 breaths/min for S\textsuperscript{2}Fusion and 7.96 for SQb Fusion. } 
				
\section{Discussion}
We presented a novel fusion approach for respiratory monitoring from multispectral videos. Our algorithm features two dedicated data fusion models that combine multichannel respiratory information to estimate the RR, and also to detect the absence of respiratory activity. For the first time, these two components are combined in a single system  together to be studied extensively. Our evaluation of videos from 30 subjects performing diverse breathing patterns showed that multisensory respiratory information combined with apneic events detection estimates RR with higher accuracy than applying single camera-based approaches\hl{, therefore demonstrating robustness during apnea events. The accuracy of the  fusion approach did not show a statistical difference between male and female subjects.}  Moreover, in contrast to several state-of-the-art methods developed for high-end thermal cameras, our multispectral video fusion pipeline enabled low-end mobile thermal cameras to satisfactorily perform  the task of respiratory monitoring.

\subsection{Respiratory rate estimation}
The SQb Fusion combined the RR estimates obtained from the RM and TA signals. The key element of the proposed fusion approach was the use of the SNR of each respiratory signal within the fusion model. Redundancy 
was important for improving the estimation of the RR,  illustrated by a significant improvement of the RMSE using fusion when compared to the results from a single spectral camera. 

We demonstrated that adopting multispectral sources provides an additional gain in accuracy. This expands the work of Pereira \textit{et al.}, who combined multiple respiratory signals from a single high-end video source to improve RR estimation \cite{pereira2018noncontact}. In combination, this enables new applications where robustness against noise and movement artifacts is essential. For example, head or torso movements can cause the temporary loss of one of the regions of interest and affect the quality of the extracted respiratory signal leading to an inaccurate estimation of the RR. We included such challenging situations into our experimental data by asking the participants in the \textit{Spontaneous Breathing} protocol to rotate their heads. Our analysis clearly showed that the SQb Fusion discarded the deteriorated, motion artifact affected estimations and emphasized those with higher SNR. However, when the movement involved both head and torso regions, as involuntarily occurred in one subject, the respiratory signals from all sources were deteriorated, resulting in a RMSE higher than 3~breaths/min. To cope with motion artifacts, dedicated computer vision methods could be further developed to quantify the motion of the whole body and use these insights as a quality assessment of the respiratory signals (e.g. by directly eliminating fusion estimates \cite{Karlen2013a}).

We analyzed the performance of the investigated RR algorithms for potential differences linked to the subjects' sex (Fig.~\ref{fig:genderAnalysis}). We found that estimating the RR using only the TA signals lead to a significantly higher RMSE for female subjects. A deeper analysis revealed that the higher estimation error on the female group was caused by an incorrect ROI localization, likely related to long hair.  No significant difference was observed for the performance of the RR algorithms based on motion signals from both NIR and FIR videos. Most importantly, no significant difference was observed  in the SQb Fusion, suggesting that the fusion algorithm was reducing the bias seen in the TA signal effectively \hlyel{and robust to sex differences}. The small size of each group (17 female and 13 male) does not allow a conclusive statistical evaluation on all algorithm components \hlyel{or demographic sub-features (such as facial hair or skin colour)}, but we believe that computer vision algorithms for physiological monitoring should systematically be tested for potential sex and other biases.

\subsection{Apnea detection}
Our apnea detector utilized \hlyel{a SVM ensemble classifier with} a set of features extracted from multiple respiratory signals as input. This enabled the detection of apneic events independently from the RR estimation and without requiring any pre-calibration steps. We demonstrated that detecting apneic events by fusing multiple respiratory signals extracted from multispectral independent videos provided higher F1 scores than adopting single spectral cameras. The importance of fusing multiple respiratory signals was underlined by the higher performance in detecting obstructive apneic events. 
During obstructive apneic events when the occlusion of the upper airways blocks air exchange, the respiratory movements of the torso were still present and could lead to a misdetection. However, the combination of the features extracted from the RM\textsubscript{NIR}, RM\textsubscript{FIR}, and TA\textsubscript{FIR} signals enabled the SVM model to accurately discriminate breathing from obstructive apneic events. Fusing the features from all the respiratory signals showed a marginal impact on the central apnea detection tasks. 
Interestingly, the RM\textsubscript{NIR} signal showed similar results to the fusion of all the features which was in contrast to its low performance in the RR estimation task (Fig.~\ref{fig:ba} b, f). A possible explanation could be that the two tasks had different requirements for signal quality. While the RR estimation task depended on high quality signals, the detection of apnea may only require processed features that were also available from low quality RM\textsubscript{NIR} signals. 

Our research and algorithm design focused on nighttime applications with monitoring of respiration in bed. This real-world scenario could easily be challenged by a body partially covered by a blanket. In Li \textit{et al.} RM is reliably extracted from NIR videos without being negatively influenced by the presence of a blanket texture \cite{Li2017}. Our experiments confirmed these findings also for the S\textsuperscript{2}Fusion and expanded the application to the task of apnea detection. In fact, we demonstrated that motion tracking applied to both NIR and FIR videos is a valid methodology to detect apneic events, even if one channel is unreliable.  

\subsection{Respiratory monitoring system}
To the best of our knowledge, the S\textsuperscript{2}Fusion is the first published fusion model that addressed the task of estimating the RR and detecting apnea in a single system. By combining the RR estimates obtained by the SQb Fusion and the apnea detection score obtained by the apnea detector, we forced the S\textsuperscript{2}Fusion to account for abnormal breathing events, \hlyel{therefore increasing the robustness of  the RR estimation } from video recordings during apneic events. The modular architecture of the fusion could enable a more granular detection of apnea events, such as the distinction between the type of apnea with a more sophisticated, non-binary classifier. 

While the presented work demonstrated the feasibility of the approach and is a significant step forward for the design of reliable non-contact respiratory monitors, it is also evident that further validation in clinical settings with patients is needed. While we have put great effort into diversifying data collection with a comparatively large number of subjects and a broadly defined protocol including spontaneous breathing, various apneic events and  confounding factors such as blanket coverage and head movements, the dataset is not fully representative of future applications. First, our subjects were healthy and breathing in the expected range of 6–22~breaths/min. Therefore, we did not investigate the performance of our algorithm for higher RR ranges as  would occur during disease or in a pediatric population. We believe that guiding healthy subjects to breathe with a RR in the range of 25-35~breaths/min can result in accentuated respiratory signs and could make the validation of any video-based algorithm for RR estimation highly biased. Furthermore, our recordings only contained simulated apneic events. The respiratory motion signals obtained by the reference measurement systems visually showed a lot of similarity to those of real patients suffering from central and obstructive sleep apnea. Nevertheless, full overnight recordings from healthy subjects and patients suffering from sleep apnea are needed to fully assess the value of non-contact respiratory monitoring for use as a telemedicine apnea assessment tool. There is significant potential for a system like ours to be implemented in overnight polysomnography screenings where cameras are already part of the standard equipment, but play a minor role in the overall assessment.

\section{Conclusion}
We presented a unique approach to monitor respiratory activity based on the fusion of multispectral videos. It explored the idea of extracting multiple respiratory signals from independent multispectral videos and strategically combining them to address RR estimation and apnea detection. 
Our findings demonstrated that fusing multiple respiratory signals from multispectral cameras increases the accuracy of respiratory activity when compared to single camera modalities. In addition, our S\textsuperscript{2}Fusion further highlighted the advantage of addressing the tasks of estimating RR and detecting apnea independently.  We experimentally demonstrated that \hl{low-cost} multispectral mobile cameras could potentially be used to monitor respiration. We, therefore, are confident that our results constitute a step towards the implementation of camera sensors to unobtrusively \hlyel{and robustly} monitor respiratory activity  both in the clinic and in remote locations.

\section{Acknowledgments}
We thank T. Meyer and R. Moser for contributing to the technical development, C. Maschio for assisting with the data collection, and J. Lim for helping to revise this manuscript. We thank the reviewers for their feedback on the manuscript. We are very grateful to all subjects for participating in our experiments.

\small

\bibliographystyle{IEEEtran}
\bibliography{refTBME20}

\includepdf[pages=-]{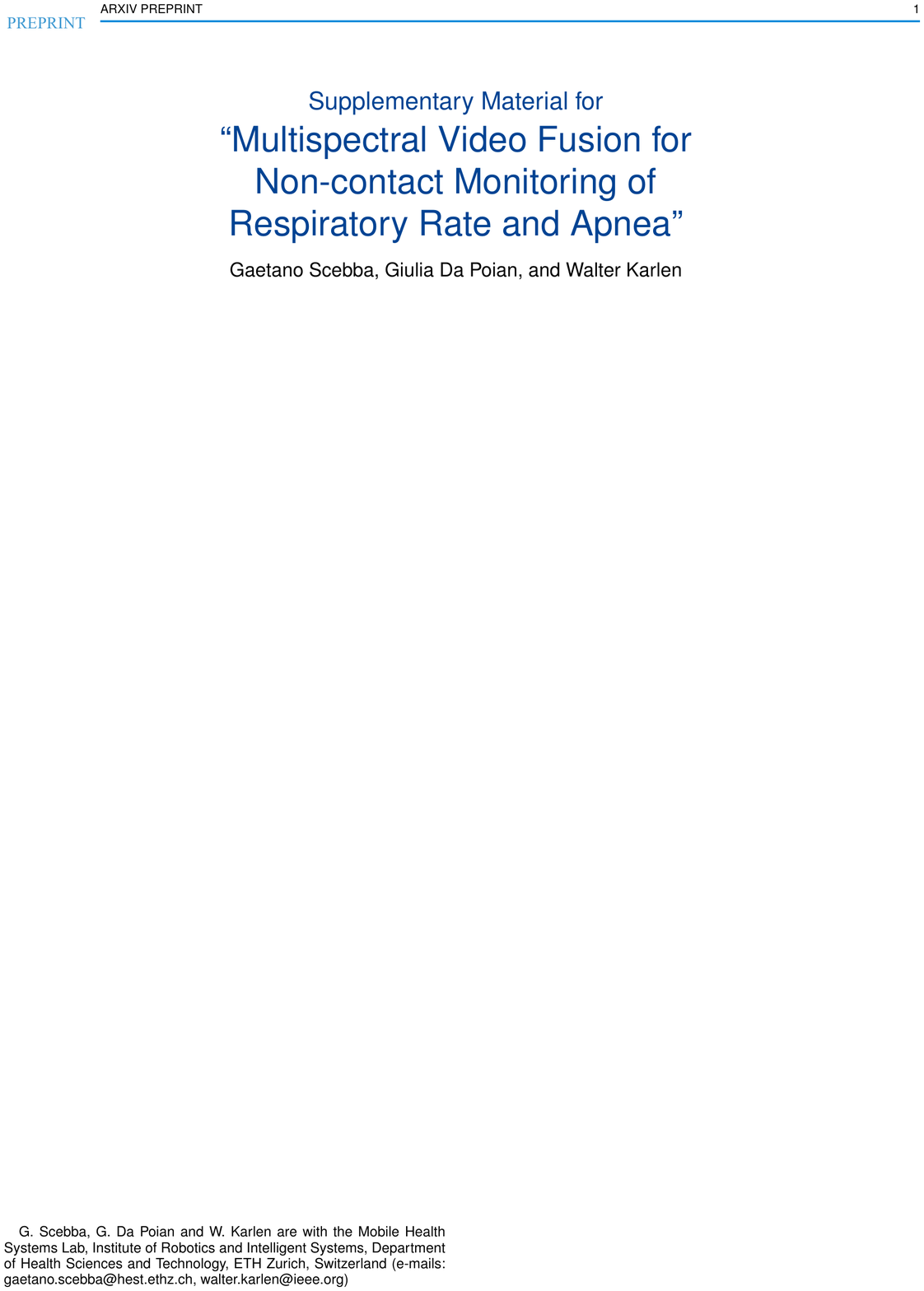}

\end{document}